\documentclass[nohyper,12pt,letterpaper]{JHEP3}

\usepackage{amsmath}
\usepackage{amsfonts}
\usepackage{amssymb}
\usepackage{graphicx}
\usepackage{epsfig}

\def\ba{\begin{eqnarray}}
\def\ea{\end{eqnarray}}
\def\be{\begin{equation}}
\def\ee{\end{equation}}
\def\d{\mathrm{d}}
\def\({\left(}
\def\){\right)}

\title{Enhanced Non-Gaussianity from Excited Initial States}
\author{R.~Holman\\
Department of Physics, Carnegie Mellon University\\ Pittsburgh PA 15213 USA\\
E-mail:  \email{rh4a@andrew.cmu.edu}}
\author{Andrew~J.~Tolley\\
Perimeter Institute for Theoretical Physics \\
31 Caroline St. N, Waterloo ON N2L 2Y5 Canada\\
E-mail: \email{atolley@perimeterinstitute.ca}}

\abstract {We use the techniques of effective field theory in an expanding universe to examine the effect of choosing an excited
inflationary initial state built over the Bunch-Davies state on the CMB bi-spectrum. We find that even for Hadamard states, there
are unexpected enhancements in the bi-spectrum for certain configurations in momentum space due to interactions of modes in the
early stages of inflation. These enhancements can be parametrically larger than the standard ones and are potentially observable
in future data. These initial state effects have a characteristic signature in $l$-space which distinguishes them from the usual
contributions, with the enhancement being most pronounced for configurations corresponding to flattened triangles for which two
momenta are collinear.}

\preprint{PI-COSMO-64}

\begin{document}

\section{\label{sec:intro}Introduction}

Inflation has become the dominant paradigm for the study of the early universe. Current data from the CMB power spectrum \cite{wmap3}, as well
as large scale structure \cite{lss} are certainly consistent with the assumption of an inflationary phase in the early universe,
while the anti-correlation between the TT and TE power spectra on superhorizon scales should soon become strong enough
statistically to serve as a ``smoking gun'' for inflation. Furthermore, the case for inflation will certainly be strengthened by
the discovery of the B-mode polarization of the CMB.

These cosmological observations provide a powerful lever arm that allows us access to physics at energy scales far beyond those
that will be probed by accelerators in the foreseeable future. Already, we can use measurements of the spectral index $n_s$ and
bounds on the ratio of tensor to scalar fluctuations $r$ to place restrictions on the form of the inflaton scalar potential, at
least within the field range corresponding to the observable 10 or so e-folds of inflation \cite{wmap3}.

The inflaton potential is only part of the story, however. The inflaton should almost certainly be viewed as an effective degree
of freedom, perhaps arising from a higher dimensional theory as the separation mode between a brane and an antibrane
\cite{braneinf}, or as one of the many axions that appears in string theory \cite{stringaxions}, amongst many other
possibilities. As such, the inflaton effective action will contain not only renormalizable terms, but also so-called irrelevant
operators which are suppressed by powers of the cutoff scale $M$. At energies higher than $M$ this effective theory breaks down
and the inflaton should not be viewed as the appropriate degree of freedom. The effect of these operators needs to be taken into
account when computing the various inflationary observables, and several works have considered these corrections \cite{irrelinf}.
Whilst these corrections might a priori have been expected to be small, they can give rise to corrections to the CMB bi-spectrum
which are larger than the standard result as shown by Creminelli \cite{Creminelli:2003iq}. This arises because the standard
contribution to the bi-spectrum is already highly suppressed by the slow roll parameters \cite{maldacena,Acquaviva:2002ud}.

Equally as important as including higher order terms in the effective action is a proper treatment of the quantum state of the
fluctuations of the inflaton about its zero mode. The standard calculations of the power spectrum are predicated on a particular
choice of quantum state, the so-called Bunch-Davies (BD) state \cite{BD}. There has been a great deal of work in recent years
focussed on understanding what might constitute reasonable modifications to this quantum state and how the data we already have
on inflationary observables from the power spectrum might constrain these modifications \cite{powerspec}.

In this work we ask the question: can we probe the nature of the initial quantum state of the inflaton through the
non-gaussianities produced via inflaton fluctuations? We will argue that statistics that probe this non-gaussianity are amazingly
sensitive to the nature of both the interactions of the inflaton and more importantly, to its initial state. In fact, in some
ways, higher correlation functions are more sensitive probes of initial state fluctuations than the power spectrum. The reason
for this is that the power spectrum is only sensitive to the interactions of the inflaton through loops which are highly
suppressed \cite{Weinberg:2005vy,Boyanovsky} (see also \cite{Sloth:2006az} for slightly different conclusions). On the other hand,
non-gaussianities can probe the inflaton's interactions directly at tree-level, which can be significant if the inflaton does not
start out in the BD vacuum state.

The simplest correlation function that probes the non-gaussian nature of inflaton statistics is the three point correlation
function, otherwise known as the bi-spectrum, of the fluctuations. The seminal calculation of the bi-spectrum was done by
Maldacena \cite{maldacena} (see also \cite{Acquaviva:2002ud}) who expanded the action for a minimally coupled inflaton scalar
coupled to gravity to third order in the gauge invariant curvature fluctuation variable $\zeta$. This is the relevant quantity
since it remains constant outside the horizon, re-emerging as a non-gaussian contribution to perturbations on scales relevant to
present day cosmology. Subsequent calculations have refined estimates of the bi-spectrum in the standard case, extended them
to other inflationary models \cite{nongauss}, and have also considered the effect of irrelevant operators \cite{Creminelli:2003iq}.

Let us briefly sketch the details of the calculations. In a cosmological setting, we need to consider the time evolution of equal
time correlation functions. Thus we choose an initial state at a (conformal) time $\eta_0$, which we take be at or near the onset
of the inflationary phase. There is considerable freedom in how to choose this state and we will discuss this issue further
below. To compute the time evolution, we work in the interaction picture where the state evolves according to \be
i\frac{d|\psi(\eta)\rangle}{d\eta}=H_I|\psi(\eta)\rangle, \ee where $H_I$ is the interaction Hamiltonian. This has the formal
solution given an initial state $|\psi(\eta_0)\rangle$ at a (finite) time $\eta=\eta_0$:  \be |\psi(\eta)\rangle =T
e^{-i\int_{\eta_0}^{\eta}H_{I}(\eta')\d\eta'}|\psi(\eta_0)\rangle. \ee We require that our effective theory be valid at
$\eta=\eta_0$. In particular, as will be discussed in Sec.~\ref{sec:eftfrw}, this means that states are not excited at momentum
scales greater than the cutoff. The initial state can be described by giving all of the correlation functions of our dynamical
degrees of freedom, e.g. the comoving curvature perturbation $\zeta(\eta)$. In particular the tree level contribution to the
equal time three point function is given by \ba && \langle \psi(\eta)|
\zeta_{\vec{k}_1}(\eta)\zeta_{\vec{k}_2}(\eta)\zeta_{\vec{k}_3}(\eta)|\psi(\eta) \rangle = \langle \psi(\eta_0)|
\zeta_{\vec{k}_1}(\eta)\zeta_{\vec{k}_2}(\eta)\zeta_{\vec{k}_3}(\eta) |\psi(\eta_0)\rangle \\ \nonumber - && i
\int_{\eta_0}^{\eta} \d\eta' \langle
\psi(\eta_0)|[\zeta_{\vec{k}_1}(\eta)\zeta_{\vec{k}_2}(\eta)\zeta_{\vec{k}_3}(\eta),H_I(\eta')]|\psi(\eta_0)\rangle + {\mathcal
O}(H_I^2).\ea Thus the total three point correlator is a sum of a contribution from any initial non-gaussianity present in the
state at the beginning of inflation, evolved forward in time with the free Hamiltonian, and the contribution that occurs from
interactions that take place between the beginning of inflation and the time of observation. The former contribution has to be
treated as arbitrary until we have a better understanding of the pre-inflationary stage. At best, we can place bounds on its
contribution based on ensuring that its backreaction is small. The second contribution, on the other hand, depends explicitly on
the interactions of the theory as well as the initial state. This contribution will be the focus of our investigations in this
work.

In practice we shall assume that the initial state is gaussian, so that all the correlation functions of operators in this state
will be completely specified by the two point Wightman function (non time-ordered in-in expectation value), which in momentum
space takes the form \be \langle \psi(\eta_0)| \zeta_{\vec{k}_1}(\eta)\zeta_{\vec{k}_2}(\eta')|\psi(\eta_0)\rangle
=(2\pi)^3\delta^3(\vec{k}_1+\vec{k}_2)G^{>}_{{k}_1}(\eta,\eta')=(2\pi)^3\delta^3(\vec{k}_1+\vec{k}_2) {\mathcal
V}_{k_1}(\eta){\mathcal V}_{k_1}^*(\eta'),\ee where ${\mathcal V}_k$ are properly normalized solutions of the linear equations of
motion for $\zeta$ in momentum space. In Sec.~\ref{sec:eftfrw} we discuss the conditions required of this two-point function so
that we can reliably trust our calculation.

\begin{figure}
    \begin{center}
     \includegraphics[width=1.0\textwidth]{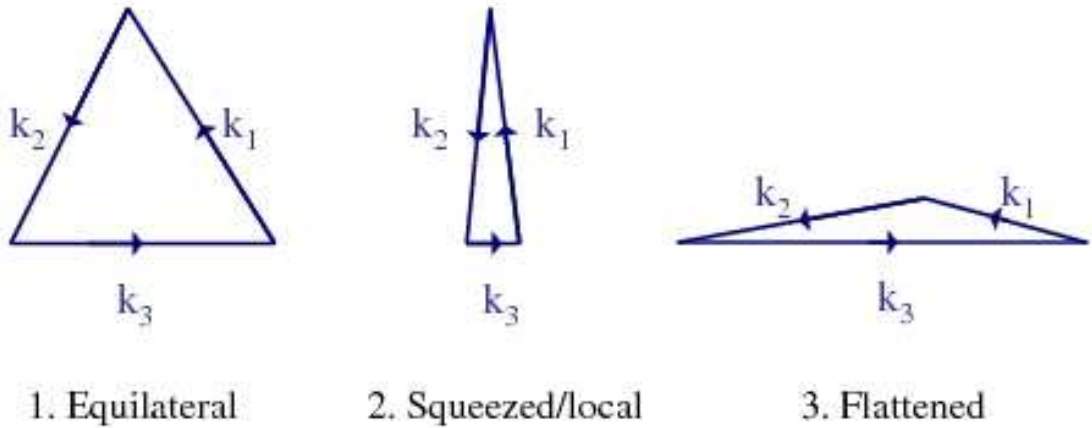}
    \end{center}
     \caption{Momentum triangles which dominate the bi-spectrum: 1.  Equilateral triangles arising from Hubble-horizon crossing.
     2. Squeezed or local triangles arising from super-Hubble evolution. 3. Flattened triangles arising from intial state effects.}
   \label{fig1}
\end{figure}

In the standard calculation using the BD vacuum, the dominant contribution to the three point function, which measures directly
three particle interactions, comes from when the modes cross the horizon. The intuitive reason for this is that at sub-horizon
scales the BD vacuum corresponds to a state of no-particles\footnote{It is true that for a static observer the BD vacuum appears
as a thermal bath of particles. However here we are using the adiabatic or WKB definition of particles appropriate to the flat
slicing of de Sitter. This is the definition that is actually most useful in the context of calculations.} (where the particles
are inflaton quanta). Thus there are no particles to interact and so no contribution to the three point function. As the modes
cross the horizon, the WKB approximation breaks down which is tantamount to the statement that particles are created. These
particles can then undergo interactions which contribute to the three point function. Once the modes are well outside the
horizon, $\zeta$, properly defined, gets frozen in nonlinearly i.e. it is conserved and further interactions become irrelevant.
In curvaton models \cite{curvaton} there can be an additional contribution from super-horizon scales because $\zeta$ is not
necessarily conserved \cite{curvatonNG}. In figure~\ref{fig1} we show the momentum triangles for which the three point function
is maximized for these two types of effects (triangles 1 and 2).

The main point of this paper is that this situation changes dramatically if the initial state is not BD. In this case there are
particles present initially which can undergo interactions. Furthermore the interactions of the inflaton are necessarily stronger
at the beginning of inflation than at the end, and so we get a second contribution to the non-gaussianity coming from
interactions in the early stages of inflation. These effects dominate the bi-spectrum for the flattened triangles (defined as
those for which two momentum are collinear) of type 3 in figure \ref{fig1} (these were called `folded' in Ref. \cite{Chen:2006nt} where a similar study was performed). To see why interactions get stronger in the past, let
us consider the prototypical example of an interacting scalar field $\phi$ on an FRW geometry
$ds^2=a^2(\eta)(-d\eta^2+d\vec{x}^2)$ with Lagrangian expressed in conformal time \be S=\int \d\eta \int  \d^3x\ a^4
\(-\frac{1}{2a^2}(\partial \phi)^2 -\sum_{n=2}^{\infty} \frac{\lambda_n}{n!M^{n-4}}\phi^n -\sum_{m=2}^{\infty}
\frac{k_m}{M^{4m-4}}\(\frac{1}{a^2}(\partial \phi)^2\)^m +\dots\) ,\ee where $M$ is the cutoff scale and the ellipsis corresponds
to additional higher derivative interactions expected on usual grounds. Upon rescaling to the canonically normalized `comoving'
field $\chi=a\phi$ we obtain\footnote{In the full gauge invariant calculation this is the familiar Mukhanov variable $v$.} \ba
S&=&\int \d\eta \int \d^3x
\(-\frac{1}{2}(\partial \chi)^2+\frac{a''}{2a}\chi^2 -\sum_{n=2}^{\infty} \frac{\lambda_n}{n!a^{n-4}M^{n-4}}\chi^n \right. \nonumber \\
&-& \left. \sum_{m=2}^{\infty} \frac{k_m}{M^{4m-4}} \frac{1}{a^{4m-4}}\((\partial \chi)^2+\frac{a'^2}{a^2}
\chi^2-2\frac{a'}{a}\chi \chi' \)^m +\dots\).\ea Note that the scaling with $a$ is the same as that with the cutoff $M$. At
subhorizon scales the terms suppressed by $a'/a$ and $a''/a$ are irrelevant. It is apparent that higher order potential
interactions with $n>4$ and higher order kinetic interactions with $m \ge 2$ necessarily grow in the past as $a \rightarrow 0$.
In other words the irrelevant operators become more relevant as modes become blue-shifted. In this case these terms will not
contribute to the three point function at tree level (the first interesting contribution is to the 4-pt from the kinetic interactions)
but as we shall see later when expanding around the background inflaton solution $\phi_0(\eta)$, we easily generate contributions
to the three point function that scale as positive powers of $1/a$. Somewhat to our surprise then, it appears that the theory with
higher derivative terms gives rise to a greater enhancement than the theory without them; the standard expectation is that
irrelevant operators would be subdominant for low energy observables such as the CMB bi-spectrum\footnote{This is a reflection of the fact that higher derivative operators are irrelevant {\em in the IR}. However, the scales that are in the IR today, relative to the cutoff scale $M$, were in the UV, or at least less in the IR at earlier times due to the expansion of the Universe.}.

The upshot of our calculation is that the bi-spectrum is {\em at least} as good a  probe of the initial state of inflaton
fluctuations as the power spectrum. The effects in both cases will depend on the Bogoliubov $\beta_k$ coefficient (see below) but
there is {\em no} extra enhancement for the case of the power spectrum. We also see that a great deal of physics can be missed by
not including the effect of irrelevant operators. Earlier works \cite{Gangui:2002qm} have also considered the effects of initial
state effects but not directly there implications for interactions. More recently in Ref.~\cite{Porrati:2004dm}, similar but less
explicit conclusions were made about the non-gaussianities. However, our approach to considering the typical corrections and in
particular the magnitude of the backreaction \cite{Porrati:2004gz} is somewhat different. The general features of the effects considered here including the precise shape dependence were pointed out in Ref.~\cite{Chen:2006nt}, although explicit estimates of the possible magnitude of these effects were not given.

In Sec.~\ref{sec:eftfrw} we discuss how effective field theories are to be construed in the context of an expanding universe, and
give a preliminary discussion of the importance of bounding backreaction. We then turn to our calculations of the bi-spectrum for
the remormalizable and higher derivative interactions of the form described above in Sec.~\ref{sec:threepoint}. We will see that
for certain triangles in momentum space, there are additional enhancements as a direct consequence of the inflaton interactions
taking place at the beginning of inflation. In Sec.~\ref{sec:measureissues} we then show that these effects have sufficient
measure, so that the enhancement survives even after the bi-spectrum is converted to spherical harmonic space. This is where the
higher derivative operators triumph over the renormalizable ones, even when the effects in both theories are calculated with the
same excited states. In Sec.~\ref{sec: hiireev} we discuss the form of the enhancements from higher irrelevant operators and
finally we conclude in Sec.~\ref{sec:concl}.

\section{\label{sec:eftfrw}Effective Field Theory in an Expanding Universe}

The usual approach to effective field theories \cite{eft} consists of the following steps. First identify applicable regime of
energies/temperatures in which we want to do physics. Next, decide what the relevant degrees of freedom in this regime should be,
as well as what symmetries their dynamics should obey. Finally write down the most general lagrangian in terms of these degrees
of freedom that incorporates the required symmetries. Once this is done, observables of the theory can be calculated in this
regime. In this calculation, a determination of how accurate the results must be (to match experimental data, say) tells us which
terms and in particular, which irrelevant operators to keep in our action.

While this approach is sufficient for standard particle physics observables, it is deficient when it comes to calculating in an
expanding universe. In this situation, we are not calculating S-matrix elements (which may not even exist in some cosmologies,
such as de Sitter \cite{wittendeS}) which depend on boundary values at asymptotic times. Instead, we need to compute the time
evolution of correlation functions which requires us to compute so-called ``in-in'' matrix elements \cite{schwingerkeldysh}, and
which incorporates the time evolution of the states, as well as the operators. In essence, we are solving a initial-value problem
rather than a boundary one.

The notion of an effective theory presupposes an energy scale $M$ such that we are constrained to do physics only below this
scale. The new twist arising from the universes' expansion is the fact that it induces a redshift of energies so that scales that
were once larger than $M$ will eventually become part of the low energy ($<M$) spectrum. In fact, we can turn this around to make
the following statement. Consider a {\em physical} momentum scale $k$ corresponding to a length scale on the CMB sky today. Then,
at a (conformal) time $\eta_0$ such that $ k\slash a(\eta_0)\sim M$ we reach the limit of validity of the effective theory for
this scale. If we want our effective theory to be valid for all of the relevant scales in the CMB sky, we must then impose a
limit on how far back in time we can trust this theory; we will let $\eta_0$ denote the earliest time at which the effective
field theory can be trusted, and we shall take this time to be the beginning of inflation, although more generally we only
require it to be the time at which the physical scales of the CMB today are of the order of the cutoff.

The scale $M$ can also infiltrate the description of the initial state of the fluctuations. The standard way this state is chosen
is by arguing that at short enough distances, the fluctuations behave as if they are in flat space, and their vacuum state would
reflect this. In practice, the $\eta\rightarrow -\infty$ limit of the solutions to the mode equation is taken, and the linear
combination of the solutions that approaches a positive energy plane wave in this limit is then chosen. The state picked by this
procedure in the case of a de Sitter universe is the Bunch-Davies (BD) state \cite{BD}; we will denote these modes via ${\cal
U}_k$ in the following.

Given that at times earlier than $\eta_0$ it may not be permissible to treat the inflaton as a well defined degree of freedom and
even if we could, we would certainly {\em not} be privy to the relevant dynamics, it does not seem reasonable to use a state
whose definition required going to arbitrarily short distances. In fact, using this state requires the radical assumption that
the description of the physics in terms of the inflaton fluctuations as a free scalar field in an FRW background is a valid one
at all scales.

A more reasonable description of the state should depend more explicitly on the domain of validity of the theory. Thus we set
initial conditions at $\eta_0$ and write
\begin{equation}
\label{eq:initialmode}
{\cal V}_k (\eta) = \alpha_k {\cal U}_k (\eta)+\beta_k {\cal U}_k^{*} (\eta),\ \left | \alpha_k \right |^2-\left | \beta_k \right |^2=1,
\end{equation}
{\it i.e.} the mode ${\cal V}_k$ is a Bogoliubov transform of the BD state. As is well known, this can be viewed as an excited
state built upon the BD state; the number density of particles of momentum $k$ is $\left | \beta_k \right |^2$. The Bogoliubov
coefficients encode information about the initial conditions satisfied by the ${\cal V}_k$ relative to the BD modes: if we
specify that
\begin{equation}
\label{eq:initiconds}
\dot{{\cal V}}_k(\eta_0) = -i\varpi_k {\cal V}_k (\eta_0),\ \ \dot{{\cal U}}_k(\eta_0) = -i\omega_k {\cal U}_k(\eta_0),
\end{equation}
then we see that
\begin{equation}
\label{eq:bogoinit}
\frac{\beta_k}{\alpha_k} = \frac{\omega_k-\varpi_k}{\omega_k+\varpi_k}.
\end{equation}

How should we fix $\beta_k$ (or equivalently, $\varpi_k$)? In Ref.~\cite{haelme} $\varpi_k$ is described in terms of a Laurent
expansion in $\omega_k\slash M$, and a new renormalization procedure has to be implemented to account for the spacelike boundary
on which the initial conditions are specified. In this work we follow the more standard approach and demand that the state
constructed from the modes ${\cal V}_k$ be Hadamard \cite{fulling}. This fixes the short-distance behavior to be the usual one,
namely that $\beta_k$ must fall of faster that $1/k^2$, but it may otherwise be arbitrary. We implement this by noticing that
since our description in terms of an effective theory breaks down at the scale $M$, we should not excite any modes with energies
higher than this. Thus we demand that $\beta_k\rightarrow 0$ for $k>Ma(\eta_0)$. Note that this assumption shields us from the
transplankian problem \cite{transplanckian} since as the universe expands and transplankian modes redshift to cisplankian scales,
they enter in their vacuum state $\beta_k=0$, and so only the subsequent cisplanckian dynamics will be important.

\subsubsection*{Backreaction}

We can put some bounds on how large the non-vanishing $\beta_k$'s can be by considering the issue of backreaction. This was
treated within the context of the standard renormalization procedure in Ref.~\cite{boyanovsky} and in
Ref.~\cite{Porrati:2004gz,haelme,schalm1} within the context of boundary renormalization. The basic issue is whether the energy
density coming from the ``particles'' of the BD vacuum contained in the initial state might overwhelm that of the inflaton zero
mode, and thus prevent inflation from occurring.

This energy density can be estimated by taking the crude model $\beta_k \sim \beta_0 e^{-k^2/(Ma(\eta_0))^2}$ and demanding that
the energy density of the nearly massless quanta of the inflaton be less than $M_{\rm pl}^2 H^2$:
\begin{equation}
\label{eq:betaconstraint} \rho\sim \frac{1}{a^4}\int \frac{\d^3 k}{\left(2\pi\right)^3}\ \left |\beta_k\right |^2 k\sim
\frac{a(\eta_0)^4}{a(\eta)^4}\left |\beta_0\right |^2 M^4\lesssim M_{\rm pl}^2 H^2\Rightarrow \left |\beta_0\right |^2\lesssim
\frac{M_{\rm pl}^2\ H^2}{M^4}.
\end{equation}
More generally we can imagine any smooth fall-off as considered in Ref.~\cite{boyanovsky}, where a fall-off of the form $\left
|\beta_k\right |^2\sim {\cal O}(k^{4+\delta})$ is assumed, and $\delta>0$ to ensure the Hadamard condition. This gives rise to
similar results (up to factors of $O(1)$), and so for the purposes of this article we shall consider the above crude model.

The stress-energy tensor also contains contributions from interactions, and these contribution are non-zero as soon as we
generate some non-gaussianities. We shall perform more explicit calculations of these contributions in
Sec.~\ref{subsec:backreaction}, but it is straightforward to estimate the bounds on these effects. On dimensional grounds, as
long as we remain in the regime of effective field theory, the largest the contribution to the energy density (expanded in powers
of $\beta_0$) can be is of order $|\beta_0|^nM^4$ for $n \ge 1$. For $|\beta_0| \le 1$ we naively expect the terms linear in
$\beta_0$ to dominate. However, any term which is odd in $\beta_0$ will also contain some powers of $e^{2ik\eta}$ inside the
momentum integrals. The same situation already arises in free theory where we get a contribution to the energy density of the
form \be \Delta \rho = -\frac{1}{a^4}\int \frac{\d^3k}{(2\pi)^3}\ k\ \beta^*_k\ e^{2ik\eta} + c.c. \, .\ee The crucial point is
that at early times, i.e. large $\eta$, the rapid oscillations of the exponential damp its contribution. For instance for the
model $\beta_k \sim \beta_0 e^{-k^2/(Ma(\eta_0))^2}$, this contribution behaves as $e^{-4\eta^2 M^2 a(\eta_0)^2}=e^{-4M^2
a(\eta_0)^2/(H a(\eta))^2}$. At early times the exponential is negligible, and once the spacetime has inflated to the point that
$a(\eta)/a(\eta_0) \approx M/H$ and the exponential is of ${\cal O}(1)$, the $1/a^4$ scaling suppresses the energy density to be
of order $H^4$ and consequently negligible. Thus even with interactions included, the backreaction to the energy density can be
no larger than $|\beta_0|^2M^4$. This is borne out by the more detailed calculation in Sec.~\ref{subsec:backreaction}.

We must also make sure that the slow roll conditions are not violated.  Since $\dot{H} =- \epsilon H^2=-\frac{1}{2M_{ m
pl}^2}\({p+\rho}\)$ and $\ddot{H} =2\epsilon \eta' H^3 =-\frac{1}{2M_{ m pl}^2}\(\dot{p}-3H(p+\rho)\)$, where $\epsilon$ and
$\eta'$ are implicitly defined slow roll parameters, then assuming $\Delta \rho \sim |\beta_0|^2 M^4$, $\Delta p \sim |\beta_0|^2
M^4$ and $\Delta \dot{p} \sim |\beta_0|^2 H M^4$ we have the bounds
\ba |\beta_0| &\le& \sqrt{\epsilon}\frac{H M_{\rm pl}}{M^2} ,\\
|\beta_0| &\le& \sqrt{\epsilon \eta'}\frac{H M_{\rm pl}}{M^2}.\ea This conclusion, also reached in \cite{schalm1}, differs strongly
from the conclusions of \cite{Porrati:2004dm,Porrati:2004gz}. That author gives more pessimistic estimates based on the order
$\beta_k$ contributions which neglect the oscillating nature of the exponentials. It is also assumed that $\Delta\dot{p} \sim M
\Delta p$ which again is not the case because of the oscillatory damping; rather we obtain $\Delta \dot{p} \sim H \Delta p$.

We see that backreaction does not necessarily force us to a small value for $\left |\beta_0\right |$. If $M\sim M_{\rm pl}$, then
$\left |\beta_0\right |\sim H\slash M_{\rm pl} \sim 10^{-6}$, while if the new physics is at scales {\em smaller} than $M_{\rm
pl}$, we can get a large value for $\left |\beta_0\right |$. In general we require $H<M$ for inflation to be described within the
regime of effective field theory, and we must also require that $M\le M_{\rm pl}$. Nevertheless for the reasonable choice $M \sim
10^{-4}M_{\rm pl}$ we can obtain $|\beta_0| \sim 1$ for $\epsilon,\eta' \sim 10^{-2}$. There are bounds on how large $\beta_k$
can be coming from direct observations of the power spectrum, but for sufficiently small or weakly $k$-dependent $\beta_k$ we can
easily evade these \cite{kinneypeiris}.

To summarize: we treat inflaton fluctuations as being described by an effective field theory valid at momenta and energies below
a scale $M$. This forces us to the notion of an earliest time $\eta_0$ at which we can trust this theory. We then argue that a
more natural set of states to use to describe inflaton fluctuations are Bogoliubov transforms of the BD state and we ensure that
they remain Hadamard by cutting off the Bogoliubov coefficient $\beta_k$ for $k>Ma(\eta_0)$. These coefficients are further
constrained by the requirement that the backreaction on the inflating background is negligible. Our assumption that the states
are Hadamard guarantees renormalizability (in the effective field theory sense). We now turn to the calculation of the
bi-spectrum using these states.
\section{\label{sec:threepoint}The Inflationary Three Point Function}

Our goal in this section is to show what impact the modifications to the quantum state of inflaton fluctuations described in
Sec.~\ref{sec:eftfrw} have on the momentum space (as well as $\ell$-space) structure of the three point function. To study
fluctuations about the FRW background, we follow Maldacena \cite{maldacena} and use the ADM foliation of the spacetime as
\begin{equation}
\label{eq:admmetric}
ds^2 = -N^2 dt^2+h_{i j}\left(dx^i+N^i dt\right)\left(dx^j+N^j dt\right),
\end{equation}
where $N$ is the lapse function and $N^i$ the shift function. The gauge invariance of the Einstein action needs to be fixed and
this can be done in one of several ways. For instance we can set the fluctuations in the inflaton field to zero $\delta \Phi=0$
and parametrize the metric fluctuations as:
\begin{equation}
\label{eq:metricflucts}
h_{i j} = \exp\left(2 H t+2 \zeta\right)\hat{h}_{i j},\ \det \hat{h} = 1.
\end{equation}
The curvature fluctuation $\zeta$ is useful since it is a gauge invariant quantity and remains constant outside the horizon. Note
that this definition of $\zeta$ differs by a sign from that often found in the literature and this is relevant to the sign of
the three point function since the bounds on $f_{NL}$ are not symmetric \cite{fnl}. For superhorizon wavelengths, gradients in
$\zeta$ can be neglected so that the $\zeta$ can be viewed as shift in the time variable. Maldacena's calculation involved using
this gauge and parametrization then expanding the Einstein action to cubic order in $\zeta$.

Another gauge that can be quite useful moves all the scalar fluctuations into the inflaton field:
\begin{equation}
\Phi\left(\vec{x}, t \right) = \phi\left(t\right) +\delta \phi\left(\vec{x},t \right),\ h_{i j} = \exp\left(2 H t\right)
\hat{h}_{i j}.
\end{equation}
In this gauge, we compute the bi-spectrum by first calculating the three point function of the fluctuations $\delta \phi$, $\langle
\delta \phi \delta \phi \delta \phi \rangle$ and then relating this to  $\langle \zeta \zeta \zeta \rangle$; it is this latter
quantity that encodes the observable non-gaussinities in the CMB. To go from $\delta \phi$ to $\zeta$ we have to take into
account the non-linear evolution of $\delta \phi$ outside the horizon which involves going to second order in the perturbations.
This gives rise to a non-linear term of the form $\langle \delta \phi \delta \phi \rangle \langle \delta \phi \delta \phi
\rangle$ to $\langle \zeta \zeta \zeta \rangle$ which serves to cancel the time dependence outside the horizon coming from the
inflaton fluctuations so that $\zeta$ remains time independent. To lowest order in slow roll parameters, which suffices for our
needs, the relation between $\delta \phi$ and $\zeta$ is
\begin{equation}
\label{eq:psizeta} \zeta =- \frac{H}{\dot{\phi}} \delta \phi + {\cal O}(\epsilon,\eta) \left(\frac{H}{\dot{\phi}} \delta
\phi\right)^2,
\end{equation}
where $\epsilon, \eta$ are the slow roll parameters. We will use this latter gauge for our calculations, and only compute the
leading order behavior in both the slow roll parameters, as well as in $H\slash M$. This means in particular that the fluctuation
modes will be those defined on de Sitter.

As discussed in Sec.~\ref{sec:eftfrw}, in a cosmological setting, the relevant expectation values are of the in-in type, as
opposed to the in-out ones used to compute S-matrix elements. There is a formalism in place for doing this
\cite{schwingerkeldysh} and it has been further elaborated in Refs.~\cite{Weinberg:2005vy,haelme,DeWitt:2003pm,Musso:2006pt}. We
use these techniques here to compute the three point function.

In the canonical version of this formalism used by Maldacena, we first construct the interacting Hamiltonian, $H_I$, in the usual
manner. The equal time tree level contribution to the three point function is then given by \be \label{eq: 3pttreecontrib}
\langle \zeta(x_1) \zeta(x_2) \zeta(x_3) \rangle = -i \int^{\eta}_{\eta_0} \d\eta' \, \langle [ \zeta(x_1) \zeta(x_2) \zeta(x_3)
,H_I(\eta')] \rangle_0; \ee using the reality condition of the equal time product this is the same as \be \label{eq:3pt} \langle
\zeta(x_1) \zeta(x_2) \zeta(x_3) \rangle = -2 \, {\mathcal Re} \(\int^{\eta}_{\eta_0} \d\eta' \, i \langle \zeta(x_1) \zeta(x_2)
\zeta(x_3) H_I(\eta')\rangle_0 \). \ee The lower limit on the integrals corresponds to the beginning of inflation $\eta_0$. This
is not an artificial cutoff, but rather it reflects our choice of a {\em gaussian} initial state. In practice, this cutoff is
crucial if we are to make sense of the integrals when initial states other than the Bunch-Davies one are chosen.

The free field correlators $\langle \zeta(x_1) \zeta(x_2) \zeta(x_3) H_I(\eta') \rangle_0$ can be computed via Wick's theorem,
where contractions are replaced by Wightman functions\footnote{This is the ``in-in'' version of the normal Wick's theorem for
time ordered products.} \be \langle {\mathcal O}_1 \zeta(x_1) {\mathcal O}_2 \zeta(x_2) {\mathcal O}_3 \rangle_0 \rightarrow
\langle \zeta(x_1)\zeta(x_2)\rangle_0 \langle {\mathcal O}_1 {\mathcal O}_2 {\mathcal O}_3 \rangle_0 + \dots \, . \ee Tadpoles
can be removed by normal ordering the interacting Hamiltonian i.e. in practice neglecting self contractions. The choice of vacuum
is then equivalent to the choice of Wightman functions \be \label{eq:wightman}\langle \zeta(x_1)\zeta(x_2)\rangle_0 =\int
\frac{\d^3 k}{(2\pi)^3}\ {\cal V}_{k}(\eta_1){\cal V}_k^*(\eta_2) e^{i\vec{k}\cdot(\vec{x}_1-\vec{x}_2)}. \ee The modes ${\cal
V}_k$ satisfy the free field equation \be {\cal V}_k''+2\frac{z'}{z}{\cal V}_k+k^2{\cal V}_k =0,\ee where $z=a\dot{\phi}/H$. We
choose them to be normalised as \be {\mathcal V}_k \rightarrow \frac{1}{\sqrt{2k}z}\(\alpha_ke^{-ik\eta}+\beta_ke^{ik\eta} \),\ee
in the limit $|k\eta| \gg 1$. The three point function will then be given by a causal time integral over products of Wightman
functions and their derivatives. We are now ready to turn to the specific calculations.

\subsection{\label{subsec:minscalar}Scalar Minimally Coupled to Gravity}

Consider now the situation treated by Maldacena, where the inflaton is a scalar field minimally coupled to gravity.  The result
contains two contributions, one from a local (in time) nonlinear field redefinition, and the second from an integral of the form
in Eq.~(\ref{eq: 3pttreecontrib}). Since local redefinitions will not contain the enhancement terms we are interested in, we can
concentrate on the non-local contribution. Maldacena \cite{maldacena} showed that up to these redefinitions the action contains
the following three point interaction for $\zeta$ (we have reexpressed his result in conformal time) \be S_{(3)}=\int \d\eta\
\d^3 x\ a^3\ M_{\rm pl}^{-2}\(\frac{\dot{\phi}}{H}\)^4H {\zeta'}^2 \partial^{-2}\zeta', \ee where $\dot{\phi}$ denotes the {\em cosmic} time
derivative of the inflaton zero mode. We use it instead of the conformal time derivative since the slow-roll conditions are
easier to express in terms of $\dot{\phi}$; $\dot{\phi}\simeq\sqrt{2 \epsilon}\ M_{\rm pl} H$. The associated interacting
Hamiltonian is $H_I=-\int \d^3 x\ a^3\ \(\frac{\dot{\phi}}{H}\)^4 H {\zeta'}^2 \partial^{-2}\zeta'$. Note that this conversion is
not entirely trivial due to the time-derivative dependence of the interaction. Substituting in the general expression Eq.~(\ref
{eq:3pt})  we obtain \ba \langle \zeta_{\vec{k}_1} \zeta_{\vec{k}_2}\zeta_{\vec{k}_3} \rangle &=& -i (2\pi)^3 \delta^3(\sum
\vec{k}_i) \(\frac{\dot{\phi}}{H}\)^4 M_{\rm pl}^{-2}H \int^0_{\eta_0} \d \eta\ a^3(\eta)\ \frac{1}{k_3^2}\
\partial_{\eta}G^{>}_{{k}_1}(0,\eta)
\partial_{\eta} G^{>}_{{k}_2}(0,\eta)\partial_{\eta} G^{>}_{{k}_3}(0,\eta) \nonumber \\&+&{\rm permutations + c.c.} \, .\ea
In writing the above expression, we have evaluated the three point function at $\eta=0$, i.e. when the modes are well outside the
Hubble-horizon. Here we see explicitly how the product of Wightman functions appears. For the BD vacuum the Wightman function is
given be \be G^{>}_{{k}_1}(\eta,\eta') =\frac{H^2}{\dot{\phi}^2}\frac{H^2}{2k^3}(1+ik\eta)(1-ik\eta')e^{-ik(\eta-\eta')}. \ee For
the first argument taken to be well after horizon crossing, we have \be
\partial_{\eta} G^{>}_{{k}_1}(0,\eta)=-\frac{H}{2k}\(\frac{H}{\dot{\phi}} \)^2 \frac{1}{a(\eta)}e^{ik\eta} \ee and so (in
the limit $\eta_0 \rightarrow -\infty$) \be \langle \zeta_{\vec{k}_1} \zeta_{\vec{k}_2}\zeta_{\vec{k}_3}\rangle =(2\pi)^3
\delta^3(\sum \vec{k}_i) \frac{1}{\prod (2k_i^3)} \frac{4 H^6}{M_{\rm pl}^2\dot{\phi}^2} \frac{\sum_{i>j} k_i^2k_j^2}{k_t}, \ee where
$k_t=k_1+k_2+k_3$, and $k_i = |\vec{k}_i |$.

Suppose we now modify the mode functions: ${\cal U}_k\rightarrow {\cal V}_k$. We get two types of corrections: one from the
modifications to the norm of the positive frequency modes $\alpha_k \, {\cal U}_{k_i}$ which just changes the overall normalisation
and shape dependences by an amount similar to that for the power spectrum. The second type of corrections which are suppressed by
powers of $\beta_k$ are more interesting. One such correction will again modify the over amplitude, but the more relevant
corrections are those that change the arguments of the oscillatory exponentials. For instance, to linear order in $\beta_k$ we
have the correction \be \Delta \langle \zeta_{\vec{k}_1} \zeta_{\vec{k}_2}\zeta_{\vec{k}_3} \rangle = -i (2\pi)^3 \delta^3(\sum
\vec{k}_i) \frac{2}{\prod (2k_i^3)} \frac{H^6}{M_{\rm pl}^2\dot{\phi}^2}  \int^0_{\eta_0} \d \eta\ \sum_j \beta^*_{k_j}
\frac{k_1^2k_2^2k_3^2}{k_j^2} e^{i \tilde{k}_{j}\eta} +{\rm c.c.} , \ee where we have defined $\tilde{k}_j=k_t-2 k_j$. Performing
the integral we have \be \label{eq:enh} \Delta \langle \zeta_{\vec{k}_1} \zeta_{\vec{k}_2}\zeta_{\vec{k}_3} \rangle = -(2\pi)^3
\delta^3(\sum \vec{k}_i) \frac{2}{\prod (2k_i^3)} \frac{H^6}{M_{\rm pl}^2\dot{\phi}^2}
  \sum_j \frac{k_1^2k_2^2k_3^2}{k_j^2\tilde{k}_j}  \beta^*_{k_j} \(1-e^{i \tilde{k}_{j}\eta_0}\) + {\rm c.c.} \, .\ee

What makes this contribution interesting are the denominators proportional to $\tilde{k}_j$ which appear. In the standard
calculation in Ref.~\cite{maldacena}, the denominator only involved $k_t=k_1+k_2+k_3$ which could never vanish unless all the
$k$'s vanished, which would then force the numerator to vanish as well. However, the mixing between positive and negative energy
BD states in our modes now allows for vanishing denominators ($\tilde{k}_j=0$) for certain {\em non-zero} values of the  momenta,
which {\em enhances} these contributions relative to those for the BD vacuum. In practice there is no divergence since the
exponential factor is unity in the $\tilde{k}_j \rightarrow 0$ limit. However, this cutoff is only relevant for
$\tilde{k}_{j}\approx 1/\eta_0$. Thus if we evaluate Eq.~(\ref{eq:enh}) for these special triangles where $\tilde{k}_j=0$, we get
\be \Delta \langle \zeta_{\vec{k}_1} \zeta_{\vec{k}_2}\zeta_{\vec{k}_3} \rangle |_{\tilde{k}_j=0} = (2\pi)^3 \delta^3(\sum
\vec{k}_i) \frac{4}{\prod (2k_i^3)} \frac{H^6}{M_{\rm pl}^2\dot{\phi}^2}
 \frac{k_1^2k_2^2k_3^2}{k_j^2} {\rm Im}(\beta_{k_j}) \eta_0,\ee
 and so relative to the standard result we get an enhancement factor \be \left .\frac{\Delta \langle \zeta_{\vec{k}_1} \zeta_{\vec{k}_2}\zeta_{\vec{k}_3} \rangle }{\langle \zeta_{\vec{k}_1} \zeta_{\vec{k}_2}\zeta_{\vec{k}_3} \rangle } \right |_{\tilde{k}_j=0} =\frac{ \frac{k_1^2k_2^2k_3^2}{k_j^2} {\rm Im}(\beta_{k_j}) k_t \eta_0}{
 \sum_s
\frac{k_1^2k_2^2k_3^2}{k_s^2}}\approx |\beta_k| |k\eta_0|= |\beta_k| \frac{k}{a(\eta_0)H}.\ee We have implicitly assumed that
${\rm Im} \beta_k \sim |\beta_k|$. Note then that although the result is suppressed by a factor of $\beta_k$, this is multiplied
by the ratio of the physical momentum at the beginning of inflation to the Hubble scale, which may be as large as $M/H$ (at least
for those modes which have been excited). In practice however, we cannot measure the full three point function but only the two
dimensional projection of it encoded into the CMB. As a result we must effectively average this result over angles which we will
do in Sec.~\ref{sec:measureissues}. On doing so we essentially lose a factor of $|k\eta_0|$ and so this result is not enhanced
but will be of the same order as similar small corrections from the local redefinitions and to the two-point function.

Nevertheless, for interactions which scale with larger powers of $1/a$ the enhancement effect can be significant and we shall
discuss these in the next section. The enhancement will occur when {\it e.g.} $k_1+k_2=k_3$. From the momentum delta function, we
also have that $|\vec{k}_1+\vec{k}_2|=\vec{k}_3$ so that only ``flattened'' triangles with two of the vectors being collinear
will be enhanced. Whether or not the effect is observable then has to do with whether these triangles have sufficient ``measure''
when the bi-spectrum is converted to $\ell$-space. We will discuss this further below.

\subsection{\label{subsec:hideriv}Higher derivative interactions}

Now let us turn to the model of higher derivative interactions discussed in \cite{Creminelli:2003iq}. It contains the following
dimension $8$ correction to the effective action for the inflaton: \be {\mathcal L}_I =\sqrt{-g}  \frac{\lambda}{8 M^4}((\nabla
\Phi)^2)^2. \ee This type of correction is regularly considered in modifications to the kinetic term of the inflaton such as in
k-inflation \cite{ArmendarizPicon:1999rj}, or DBI inflation \cite{Silverstein:2003hf}. Since we do not want to be wedded to a specific scenario, for the purposes of the present discussion
we shall concentrate on standard slow-roll inflation; however, it is straightforward to extend the arguments to these more
general cases.

Before looking at its contribution to the three point function, we shall make some cautionary remarks about the tree level contribution
to the two point function coming from this interaction. Expanding $\Phi=\phi(\eta)+\delta \phi$ to second order in $\delta \phi$
and then converting from $\delta \phi$ to $\zeta$, this term contributes the following term to the Hamiltonian \ba
H_{I}^{(2)}&=& \frac{\lambda}{4 M^4} \int \d^3 x a^2 \dot{\phi}^2 \( 3{\delta \phi'}^2-(\partial_i \delta \phi)^2 \) \nonumber\\
&=& \frac{\lambda}{4 M^4} \int \d^3 x a^2 \frac{\dot{\phi}^4}{H^2} \( 3(\zeta'+ar\zeta)^2-(\partial_i \zeta)^2 \) , \ea where
$r=\frac{d\ln(\dot{\phi}/H)}{dt}$. Since $r$ is both slow roll suppressed and its contribution to the action subdominant at
subhorizon scales we can ignore it for the present discussion. Since this term is quadratic in the fields, we could, in principle, simply incorporate it into the
free Hamiltonian. On the other hand, we could also treat this as an interaction, just as for the three point function. Doing this would yield the following correction to the $\zeta$ equal time two point function evaluated at late times \be \label{eq:twoptcorrect}
\Delta \langle \zeta(\vec{x}_1)\zeta(\vec{x}_2) \rangle = -\frac{i\lambda}{2M^4} \int_{\eta_0}^{0} \d {\eta} a^2({\eta})
\frac{\dot{\phi}^4}{H^2} \left( 3 (\partial_{{\eta}} G_{k}^{>}(0,{\eta}))^2-k^2 G_{k}^{>}(0,{\eta})^2 \right)+{\rm c.c.}. \ee On
considering the terms linear in $\beta_k$, this appears to give similar enhancements as described in the previous section. In
this case the enhancements will not get washed out by averaging over angles. However, this is clearly a fake since had we
incorporated this term into the free Hamiltonian we would not have seen any such affect. The resolution is that this term can be
removed by a renormalization of the initial vacuum choice $\beta_k \rightarrow \beta_k+\lambda \delta \beta_k$.

To see where the problem arises we can always resum two point vertices by absorbing them into the definition of the free
Hamiltonian. On doing so, the Wightman function will be constructed from free modes satisfying an equation of the form \be {\cal
V}_k^{\prime \prime}+2\frac{z'}{z} {\cal V}_k +c_s^2 k^2 {\cal V}_k =0 , \ee where \be c_s^2=\frac{(1-\lambda
\dot{\phi}^2/(2M^2))}{(1-\lambda \dot{\phi}^2/(3M^2))} .\ee Here we see the modified sound speed characteristic of models such as
k-inflation \cite{ArmendarizPicon:1999rj} and DBI inflation \cite{Silverstein:2003hf} that make use of these higher derivative
operators. At subhorizon scales the WKB approximation is valid, and so can write the general solution as \be {\cal V}_k
=\frac{1}{\sqrt{2c_sk} z}\left(\alpha_k \exp(-ic_s |k|\eta )+\beta_k \exp(+i c_s |k| \eta )\right). \ee   We can now see the
reason for the apparently large corrections to the power spectrum. If we expand these modes out to ${\cal O}(\lambda)$ we get \be
\exp(-ic_s |k|\eta)=\exp(-i|k|\eta) (1-i |k| \eta \frac{\lambda \dot{\phi}^2}{2M^4} + O(\lambda^2)), \ee which is  a poor
expansion for sufficiently large $|k \eta|$. From now on we shall work with the resummed mode functions ${\cal V}_k$ which are
well behaved at subhorizon scales, which has the effect of removing the correction in Eq.~(\ref{eq:twoptcorrect}); since the
correction to the three point function computed below is already ${\cal O}(\lambda)$, we can omit these corrections to the modes.

Having taking care of these issues, we can now turn to the calculation of the three point function in the presence of these
higher derivative interactions. Expanding the interaction Lagrangian to third order in the fluctuations $\delta \phi$ and then
converting to $\zeta$, we find \ba { H}^{(3)}_I&=& -\int \d^3x a \frac{\lambda \dot{\phi}}{2 M^4} \delta \phi' \left({\delta
\phi'}^2-\left(\partial_i \delta \phi\right)^2\right),\nonumber \\
&=&-\int \d^3x a \frac{\lambda \dot{\phi}^4}{2 H^3 M^4} \zeta' \left({\zeta'}^2-\left(\partial_i \zeta\right)^2\right),
 \ea
where again we have neglected the $r$ term. Note that although the original operator was dimension 8, this operator is really
dimension 6 in terms of $\delta \phi$, but it is additionally suppressed by $\sqrt{\epsilon} \( HM_{\rm pl}\slash M^4\)$.

We can now compute the corrections of interest to the three point function. We compute to lowest order in $\lambda$, $H\slash M$,
and $\epsilon$ to find \ba \label{eq:3ptcorrehider} \langle \zeta_{k_1}  \zeta_{k_2}  \zeta_{k_3} \rangle &=& \left(2\pi\right)^3
\delta^{(3)} \left(\sum_i \vec{k}_i\right) {\cal A}\left(\vec{k}_1, \vec{k}_2,
\vec{k}_3\right)\nonumber \\
{\cal A}\left(\vec{k}_1, \vec{k}_2, \vec{k}_3\right) &=&i \frac{ \lambda\dot{\phi}^4}{2H^3 M^4}\int_{\eta_0}^0 \d\eta\ a
\left( \partial_{\eta} G_{k_1}^> (0, \eta) \partial_{\eta} G_{k_2}^> (0, \eta)\partial_{\eta} G_{k_3}^> (0, \eta)+\right. \nonumber \\
 &+& \left.(\vec{k}_1\cdot \vec{k}_2) (G_{k_1}^> (0, \eta) G_{k_2}^> (0, \eta)\partial_{\eta} G_{k_3}^> (0, \eta))+{\rm perms.}
\right) +{\rm c.c.}. \ea The corrections we want are again those linear in $\beta_k$. These give rise to the following
contribution to the three point function $\zeta_k$. \be \label{eq: hiderivcurv} \Delta \langle \zeta_{\vec{k}_1}
\zeta_{\vec{k}_3} \zeta_{\vec{k}_3} \rangle = \left(2\pi\right)^3 \delta^{(3)} \left(\sum_i \vec{k}_i\right) \frac{1}{\prod
\left(2 k_i^3\right)} \frac{i \lambda H^8}{\dot{\phi}^2 M^4}\sum_j \beta^*_{{k}_j}  \int_{\eta_0}^{0} \d \eta e^{i\tilde{k}_j
\eta}S_j(k_1,k_2,k_3,\eta)+{\rm c.c.},\ee where \ba &&S_j(k_1,k_2,k_3,\eta)=-k_t\(\Pi_i \tilde{k}_i\) + \\ &&
(-\eta)\(k_j^4(k_{j+1}+k_{j+2}) +k_{j+1}k_{j+2}(k_{j+1}-k_{j+2})^2(k_{j+1}+k_{j+2})-k_j^3(k_{j+1}^2+k_{j+2}^2) \right. \nonumber
\\ \nonumber && \left. - k_j^2(k_{j+1}+k_{j+2})(k_{j+1}^2+k_{j+1}k_{j+2}+k_{j+2}^2)\) +(-\eta)^2 \tilde{k}_j \( \Pi_i k_i \)\( k_t^2-4k_{j+1}k_{j+2}\) ,\ea where $j$ is defined modulo 3.
As in Sec.~\ref{subsec:minscalar}, we see that these results are enhanced for the flattened triangles where two of the vectors
are collinear so that $\tilde{k}_{i}$ vanishes for some $i$. Note that the first and third terms in the $\eta$ expansion of $S_j$
vanish and so the dominant contribution evaluated on a give flattened triangle $\tilde{k}_j=0$ is \ba \left. \Delta \langle
\zeta_{\vec{k}_1} \zeta_{\vec{k}_3} \zeta_{\vec{k}_3} \rangle \right|_{\tilde{k}_j=0} &=& -\left(2\pi\right)^3 \delta^{(3)}
\left(\sum_i \vec{k}_i\right) \frac{1}{\prod \left(2 k_i^3\right)} \frac{\lambda H^8}{\dot{\phi}^2 M^4} 2 {\mathcal Re}(\beta_{{k}_j}) \\
\nonumber && \times \frac{\eta_0^2}{2}\( 4k_{j+1}k_{j+2} (k_{j+1}+k_{j+2})(k_{j+1}^2+k_{j+2}^2+k_{j+1}k_{j+2})\).\ea For these
specific triangles the enhancement factor relative to the BD contribution is \be \left. \frac{\Delta \langle
\zeta^3\rangle}{\langle \zeta^3\rangle} \right|_{\rm flattened} \approx |\beta_k| |k\eta_0|^2, \ee where we have assumed
${\mathcal Re}(\beta_k) \approx |\beta_k|$. Again as will be explained in the next section on going to $l$ space we effectively
lose one factor of $|k\eta_0|$, but one enhancement factor remains. Thus we can give an order of magnitude estimate for $f_{NL}$
for these triangles to be \be \left . f_{NL}\right |_{\rm flattened}\sim \frac{\dot{\phi}^2}{M^4} |\beta_k| \(\frac{k}{a(\eta_0)
H}\).\ee For backreaction to be under control, the largest reasonable value for $|\beta_k|$ is $\sqrt{\epsilon \eta'} HM_{\rm
pl}/M^2$ while for the effective field theory to be valid the largest value allowed for $k/{a(\eta_0)}$ is $M$. Thus the maximum
expected contribution to $f_{NL}$ is \be f_{NL}|_{\rm flattened} \sim \sqrt{\epsilon \eta'}\frac{\dot{\phi}^2}{M^4}\frac{HM_{\rm
pl}}{M^2}\(\frac{M}{H}\) \sim \epsilon \sqrt{\epsilon \eta'}\( \frac{H}{M_{\rm pl}}\)^2 \(\frac{M_{\rm pl}}{M}\)^5. \ee

We see then that for reasonable values of these parameters $\epsilon,\eta' \sim 10^{-2}$, $H/M_{\rm pl}\sim 10^{-6}$ we get \be
\left .f_{NL}\right |_{\rm flattened} \sim \(6.3 \times 10^{-4} \, \frac{M_{\rm pl}}{M}\)^5, \ee which is in the range $1$ to
$100$ for $M$ in the range $6.3 \times 10^{-4}M_{\rm pl}$ to $2.5 \times 10^{-4}M_{ \rm pl}$. In pushing into the limit $f_{NL}
\sim 100$ we have $|\beta_k| \sim 1.6 \times 10^{-1}$ which may or may not be already observable in the power spectrum depending
on the precise $k$ dependence of the $\beta_k$. In practice these bounds are overly restrictive since we know that models such as
DBI inflation do not have to satisfy the usual slow roll restrictions, and in fact the more general case was considered in Ref.~\cite{Chen:2006nt}. In these models the usual contribution to the
non-gaussianity may be large, and so this specific contribution will be further enhanced. In general the enhancement factor for
these specific triangles relative to the usual contribution is bounded by \be \frac{{\left . f_{NL} \right |_{\rm
flattened}}}{{\left . f_{NL} \right |_{\rm usual}}} \sim |\beta_k| \frac{M}{H} \sim 100 \sqrt{|\beta_k|}. \ee From an
observational point of view, one might think that since only specific triangles give the enhancement, this effect might be buried
in the noise. However, consider  the $l=500$ total modes which have been observed by WMAP with signal to noise greater than 1. In
total, the bi-spectrum is made of $(500)^3$ points (modulo symmetry factors). Of these, the total number that satisfy the
triangle inequality are roughly $(500)^2$ and so are down by a factor of $500$. This would then imply that the signal to noise
coming from these will thus be down by a factor of $1/\sqrt{500}$. Thus, roughly 5 percent of the signal will come from these
modes, and so to a crude approximation $\sim 10^2 \sqrt{|\beta_k|}$ enhancement will effectively only amount to a $\sim 10
\sqrt{|\beta_k|}$ enhancement. In practice it is necessary to reanalyse the data with an appropriate template along the lines of
Ref.~\cite{fnl}, and this will improve the constraints on this contribution.

\subsection{Backreaction from higher derivative interactions}
\label{subsec:backreaction}

We have already given a simple estimate for the absence of backreaction based on computing the expectation value of the free
field stress energy. However, as soon as we choose a non-standard initial state, and a non-zero gaussianity develops we will also
generate additional cubic and higher order contributions to the stress energy. As usual we can infer these via the in-in
formalism \cite{schwingerkeldysh} so that the expectation value of the stress energy at any time $\eta$ is given be \be \langle
T_{\mu\nu}(\eta,x) \rangle = \langle \bar{T}e^{i\int_{\eta_0}^{\eta} \d \eta' H_I} T_{\mu\nu}(\eta,x)Te^{-i\int_{\eta_0}^{\eta}
\d \eta' H_I} \rangle_0,
 \ee
 which gives to first order in the interaction
 \be
 \label{eq:stresscorrection}
\langle T_{\mu\nu}(\eta,x) \rangle = -2 {\mathcal Re} \( i \int_{\eta_0}^{\eta} \d \eta' \langle T_{\mu\nu}(\eta,x)
H_{I}\rangle_0 \),
 \ee
where it is important to not that the $T_{\mu\nu}$ inserted on the right hand side is defined in the interacting representation.
Thus in particular a contribution to $T_{\mu\nu}$ cubic in the interaction muliplied by the cubic vertex in the Hamiltonian will
give a
 non-zero expectation value. If we choose the Bunch-Davies vacuum then for the usual reasons we can safely ignore this backreaction
 contribution since the vacuum is de Sitter invariant and it will at most contribute to a renormalization of the zero point of
 the potential. If, however, we choose instead an excited yet still gaussian initial state, the early non-gaussianity that is generated may cause a problem.

 Adding to the action the interaction $\frac{\lambda}{8M^4} (\partial \phi)^4$ gives the following
 contribution to the stress energy (in the Heisenberg rep.)
 \be
\Delta T^H_{\mu\nu} = -\frac{\lambda}{2M^4} (\partial \phi)^2 \partial_{\mu}\phi\partial_{\nu}\phi+\frac{\lambda}{8M^4}(\partial
\phi)^4 g_{\mu\nu}.
 \ee
 In particular, expanding to cubic order around an inflating solution $\phi(\eta)$ (and neglecting backreaction of the metric which is a reasonable approximation at sub-Hubble scales)
  we get the following cubic contribution to the energy density
\be \Delta_{(3)} \rho^H = \frac{\lambda \dot{\phi}}{2M^4a^3} \delta{\phi'}\(3\delta {\phi'}^2 -(\partial_i \delta \phi)^2\) .\ee
On converting to the interacting representation we unsurprisingly obtain the interacting Hamiltonian density\footnote{To get from
one to the other one first writes the Heisenberg expression in terms of $\phi$ and $\pi$, and then replace $\pi$ by its free
field value $\pi=a^3 \dot{\phi}$.} \be \Delta_{(3)} \rho = \frac{\lambda \dot{\phi}}{2M^4a^3} \delta{\phi'}\(\delta {\phi'}^2
-(\partial_i \delta \phi)^2\) .\ee
 In any case rather than performing the full calculation we
may simply note that at sub-Hubble scales $\delta \phi'$ and $\partial_i \delta \phi$ scale the same way so the effective
contributions to the pressure and stress energy take the form (where we have defined $\chi=a\delta \phi$)\footnote{In fact the
final answer is smaller since the Lorentz invariant combination $\delta \phi'^2-(\partial_i \delta \phi)^2$ gives an additional
$H/M$ suppression.} \ba \Delta_{(3)} \rho &\approx& \frac{\lambda^2 \dot{\phi}^2}{M^8 a^6(\eta)} \int_{\eta_0}^{\eta}\d\eta' \int
\d^3 x' \frac{1}{a^2(\eta')} \langle
{\chi'}^3(\eta,x){\chi'}^3(\eta',x') \rangle_0  \\
&\approx &\frac{\lambda^2 \dot{\phi}^2}{M^8 a^6(\eta)} \int_{\eta_0}^{\eta}\d\eta' \int \d^3 x' \frac{1}{a^2(\eta')}
\(\partial_{\eta}\partial_{\eta'}H(x',\eta-\eta')\)^3,\ea 
where 
\ba H(x',\eta-\eta')&=&\int \frac{\d^3k}{(2\pi)^3}
\frac{1}{2|k|}e^{-i\vec{k}.\vec{x}'}\(|\alpha_k|^2e^{-i|k|(\eta-\eta')} \right . \nonumber \\ 
&+& \left. \alpha_k^*\beta_k e^{i|k|(\eta+\eta')} +\alpha_k\beta^*_k e^{-i|k|(\eta+\eta')}+|\beta_k|^2 e^{i|k|(\eta-\eta')} \) .
\ea
This gives at worst a contribution of order \be \Delta_{(3)}\rho \approx
\frac{\lambda^2 \dot{\phi}^2}{M^8a^6}{\mathcal Re} \int^{\eta}_{\eta_0} \d\eta'\frac{1}{a^2(\eta')} \int \frac{\d^3
k_1}{(2\pi)^3} \int \frac{\d^3 k_2}{(2\pi)^3} k_1k_2|\vec{k}_1+\vec{k}_2|\beta^*_{k_1} e^{-2 i k_1 \eta}
e^{-i(-k_1+k_2+|\vec{k}_1+\vec{k}_2|)(\eta-\eta')},\ee to first order in $\beta_k$. As explained in Sec.~\ref{sec:eftfrw} the
presence of the oscillatory factor $e^{\pm 2i k_1 \eta}$ washes out the early time contribution to this integral. The leading
contribution is then the ${O}(\beta_k^2)$ contribution which is bounded by \be \Delta \rho_{(3)} \lesssim |\beta_0|^2
\dot{\phi}^2 .\ee To obtain this estimate we can simply count the positive powers of $k$ and $\eta$ in the integral and replace
them with their maximum possible values $a(\eta_0)M$ and $\eta_0$ respectively, and then dividing by one factor of $|k\eta_0|$
from the angular integration. Note that since we also require $\dot{\phi}^2<M^4$ for the validity of the effective field theory
this contribution is necessarily smaller than the free field contribution.

Similarly at quartic order we have in the Heisenberg representation \be \Delta_{(4)} \rho^H = \frac{\lambda}{8M^4a^4} \(3\delta
{\phi'}^4-2 (\delta \phi')^2(\partial_i \delta \phi)^2 -(\partial_i \delta \phi)^4\) . \ee This gives a contribution of the form
\ba \Delta_{(4)} \rho &\approx& \frac{\lambda^2}{M^8 a^8(\eta)} \int_{\eta_0}^{\eta}\d\eta' \int \d^3 x' \frac{1}{a^4(\eta')}
\langle
{\chi'}^4(\eta,x){\chi'}^4(\eta',x') \rangle_0  \\
&\approx &\frac{\lambda^2}{M^8 a^8(\eta)} \int_{\eta_0}^{\eta}\d\eta' \int \d^3 x' \frac{1}{a^4(\eta')}
\(\partial_{\eta}\partial_{\eta'}H(x',\eta-\eta')\)^4.\ea As before the dominant contribution is bounded by \be \Delta \rho_{(4)}
\lesssim |\beta_0|^2 M^4. \ee According to the general argument given in Sec.~\ref{sec:eftfrw} higher order contributions will
never give a contribution larger than this as long as we remain in the validity of the effective theory.

\section{Measure issues and the l-space bi-spectrum}

\label{sec:measureissues}

The results of the previous section show that for flattened triangles the three point function can be considerably enhanced
relative to the usual case. However, we do not actually measure the three dimensional three point function directly in the CMB.
Rather, we measure its two dimensional projection as encoded in the $a_{l m}$'s.

The relevant question then is to what extent these enhancements can be seen in the CMB l-space three point function. Once the
temperature fluctuations have been decomposed into spherical harmonics, \be \frac{\Delta T}{T}(\hat{n})
=\sum_{lm}a_{lm}Y_{lm}(\hat{n}), \ee we can construct the angular averaged bi-spectrum \cite{Babich:2004gb} \be B(l_1,l_2,l_3) =
\sum_{m_1,m_2,m_3} \(
\begin{array}{ccc}
l_1 & l_2 & l_3 \\
m_1 & m_2 & m_3
\end{array}
 \)
 \langle a_{l_1m_1}a_{l_2m_2}a_{l_3m_3} \rangle.
\ee Expressing the three point function for $\zeta_k$ as \be \langle \zeta_{\vec{k}_1} \zeta_{\vec{k}_2} \zeta_{\vec{k}_3}
\rangle = (2\pi)^3 \delta(\sum_i \vec{k}_i)\ {\mathcal A}(\vec{k}_1,\vec{k}_2,\vec{k}_3), \ee the three point function of the
$a_{lm}$ is given by (in this section we will make use of the unit vectors defined via $\vec{k}_i=|k_i|\hat{n}_i$). \ba
 \langle a_{l_1m_1}a_{l_2m_2}a_{l_3m_3} \rangle &=& (4\pi)^3 i^{l_1+l_2+l_3} \int \frac{\d^3\vec{k}_1}{(2\pi)^3}
 \frac{\d^3\vec{k}_2}{(2\pi)^3} \frac{\d^3\vec{k}_3}{(2\pi)^3}\ Y^*_{l_1m_1}(\hat{n}_1)Y^*_{l_2m_2}(\hat{n}_2)Y^*_{l_3m_3}(\hat{n}_3) \nonumber\\
 && (2\pi)^3 \delta^{(3)}(\sum_i \vec{k}_i)\ {\mathcal A}(\vec{k}_1,\vec{k}_2,\vec{k}_3) \Delta^T_{l_1}(k_1)\Delta^T_{l_2}(k_{2})\Delta^T_{l_3}(k_3),
\ea where $\Delta_l^T(k)$ are the radiation transfer functions. We are interested in a specific combination to ${\mathcal A}$
that comes from flattened triangles. The contributions to the three point function linear in $\beta_k$ can be written in the form
\be \label{eq:amplitude}{\mathcal A}(\vec{k}_1,\vec{k}_2,\vec{k}_3) = \sum_j \beta^*_{k_j}(ik_t)^{s+1} \int^0_{\eta_0} \d\eta
(-\eta)^{s}e^{i\tilde{k}_j\eta} {\mathcal D}(k_j,k_{j+1},k_{j+2})+c.c. , \ee where $j$ is defined modulo 3. Let us consider the
properties of the integral \be I(\tilde{k}_i)= \int^0_{\eta_0} \d\eta (-\eta)^{s}e^{i\tilde{k}_i\eta}.\ee For
$|\tilde{k}_i\eta_0| \gg 1$ the oscillatory nature of the integrand heavily damps the power of $\eta$ and as a result the
integral scales as $(1/i\tilde{k}_i)^{s+1}$. On the other hand for $|\tilde{k}_i\eta_0| \ll 1$ the exponential is irrelevant and
the integral gives $I(\tilde{k}_i)=(-\eta_0)^{s+1}/(s+1)$. We are interested in effects coming from sub-Horizon scales where
$|k_i\eta_0| \gg 1$, and so the dominant contribution to the integral will come from the regime where the triangles are flattened
and $\tilde{k}_i \approx 0$ so that the integral is given by its maximum value $I(\tilde{k}_i)=(-\eta_0)^{s+1}/(s+1)$. However,
this will only occur for a special range of angles, essentially for \be \hat{n}_i . \hat{n}_j+1 \ll \frac{(k_i-k_j)^2}{k_i
k_j}.\ee

To perform the integrals, let us first integrate out $\vec{k}_{j+2}$ via the delta function, and then integrate over
$\hat{n}_i$ using $\hat{n}_{i+1}$ as a reference axis. On doing so we get \ba &&\langle a_{l_1m_1}a_{l_2m_2}a_{l_3m_3} \rangle \nonumber 
= \sum_j (4\pi)^3 i^{l_1+l_2+l_3} \int \frac{\d k_j k_j^2}{(2\pi)^3} \int \frac{\d k_{j+1} k_{j+1}^2}{(2\pi)^3} \beta^*_j \int
\d^2\hat{n}_{j+1} \int \d^2\hat{n_j} \, \\ \nonumber
&& Y^*_{l_jm_j}(\hat{n}_j)Y^*_{l_{j+1}m_{j+1}}(\hat{n}_{j+1})Y^*_{l_{j+2}m_{j+2}}(\hat{n}_{j+2}) 
(ik_t)^{s+1} I(\tilde{k}_j) {\mathcal D}({k}_j,{k}_{j+1},|\vec{k}_j-\vec{k}_{j+1}|) \times
\\ 
&& 
\Delta^T_{l_j}(k_j)\Delta^T_{l_{j+1}}(k_{j+1})\Delta^T_{l_{j+2}}(|\vec{k}_{j}-\vec{k}_{j+1}|) + \dots \, , \ea where the ellipsis
represents the terms arising from the complex conjugate in Eq.~(\ref{eq:amplitude}). At this point we can make the approximation
that since the integral $I(\tilde{k}_j)$ is so strongly peaked at $\tilde{k}_j=0$ we may replace the product of spherical
harmonics with their values for the flattened triangles set by $\tilde{k}_j=0$, namely \be
Y^*_{l_jm_j}(\hat{n}_j)Y^*_{l_{j+1}m_{j+1}}(\hat{n}_{j+1})Y^*_{l_{j+2}m_{j+2}}(\hat{n}_{j+2}) \rightarrow
Y^*_{l_jm_j}(-\hat{n}_{j+1})Y^*_{l_{j+1}m_{j+1}}(\hat{n}_{j+1})Y^*_{l_{j+2}m_{j+2}}(\hat{n}_{j+1}). \ee Similarly we may replace
\be \Delta^T_{l_{j+2}}(|\vec{k}_{j}-\vec{k}_{j+1}|) \rightarrow \Delta^T_{l_{j+2}}(|{k}_{j}-{k}_{j+1}|). \ee In this
approximation, the integral $I(\tilde{k}_j)$ gets directly integrated over angles \be \int \d^2\hat{n}_j I(\tilde{k}_j) = 2\pi
\int^0_{\eta_0} \d\eta (-\eta)^{s}\frac{1}{k_jk_{j+1}\eta^2}\(e^{2ik_{j+1}\eta}\(1-i(k_j+k_{j+1})\eta\) -
\(1-i(k_j-k_{j+1})\eta\) \),\ee where we have assumed $k_j \ge k_{j+1}$ as is necessary for the existence of this triangle. In
the limit $|k_s\eta_0| \gg 1$ this integral is given approximately by \be \int \d^2\hat{n}_j I(\tilde{k}_j) \approx \frac{2\pi i
(k_j-k_{j+1})}{sk_jk_{j+1}}(- \eta_0)^{s}. \ee 
So we have 
\ba &&\langle a_{l_1m_1}a_{l_2m_2}a_{l_3m_3} \rangle \nonumber
\approx \sum_j  (4\pi)^3 i^{l_1+l_2+l_3} \int \frac{\d k_j k_j^2}{(2\pi)^3} \int_{k_{j+1}<k_j} \frac{\d k_{j+1}
k_{j+1}^2}{(2\pi)^3}
 \int \d^2\hat{n}_{j+1} \, \\ \nonumber
&& Y^*_{l_jm_j}(-\hat{n}_{j+1})Y^*_{l_{j+1}m_{j+1}}(\hat{n}_{j+1})Y^*_{l_{j+2}m_{j+2}}(\hat{n}_{j+1})  \beta^*_{k_j} \frac{2\pi k_t (k_{j+1}-k_{j})}{sk_jk_{j+1}} (-ik_t\eta_0)^{s} \times
 \\ 
&&{\mathcal D}({k}_j,{k}_{j+1},|k_j-k_{j+1}|)
\Delta^T_{l_j}(k_j)\Delta^T_{l_{j+1}}(k_{j+1})\Delta^T_{l_{j+2}}(|k_j-k_{j+1}|) + {\dots}. \ea 
Using the fact that
$Y_{lm}(-\hat{n})=(-1)^l Y_{lm}(\hat{n})$ we get \ba &&\langle a_{l_1m_1}a_{l_2m_2}a_{l_3m_3} \rangle \approx \sum_j (4\pi)^3
i^{l_1+l_2+l_3} (-1)^{l_1} \int \frac{\d k_j k_j^2}{(2\pi)^3} \int_{k_{j+1}<k_j} \frac{\d k_{j+1} k_{j+1}^2}{(2\pi)^3}
\beta^*_{k_j}\, {\mathcal G}_{l_1,l_2,l_3}^{m_1,m_2,m_3} \nonumber \\ &&  \frac{2\pi k_t(k_{j+1}-k_{j})}{sk_jk_{j+1}}
(-ik_t\eta_0)^{s} {\mathcal D}({k}_j,{k}_{j+1},|k_j-k_{j+1}|)
\Delta^T_{l_1}(k_j)\Delta^T_{l_2}(k_{j+1})\Delta^T_{l_3}(|k_j-k_{j+1}|) + {\dots }, \nonumber \ea where ${\mathcal
G}_{l_1,l_2,l_3}^{m_1,m_2,m_3}$ is the Gaunt integral \cite{Babich:2004gb}. Finally then, these contributions to the bi-spectrum
take the form \ba &&B(l_1,l_2,l_3) = \\ \nonumber &&\sum_j \sqrt{\frac{(2l_1+1)(2l_2+1)(2l_3+1)}{4\pi}} \(
\begin{array}{ccc}
l_1 & l_2 & l_3 \\
0 & 0 & 0
\end{array}
 \) (4\pi)^3 i^{l_1+l_2+l_3} (-1)^{l_1} \int \frac{\d k_j k_j^2}{(2\pi)^3} \int_{k_{j+1}<k_j} \frac{\d k_{j+1} k_{j+1}^2}{(2\pi)^3}
  \\
&&  \beta^*_{k_j}\frac{2\pi k_t (k_{j+1}-k_{j})}{sk_jk_{j+1}} (-ik_t\eta_0)^{s} {\mathcal D}({k}_j,{k}_{j+1},|k_j-k_{j+1}|)
\Delta^T_{l_j}(k_j)\Delta^T_{l_{j+1}}(k_{j+1})\Delta^T_{l_{j+2}}(|k_j-k_{j+1}|) + {\rm c.c.}. \nonumber \ea The crucial point to
note is that whilst the original integral naively scales as $|k_t \eta_0|^{s+1}$, the above bi-spectrum only contains a factor
$|k_t\eta_0|^s$. In short, in performing the 2-d projection we essentially have to smooth the triangles over some finite
resolution of solid angles. This softens the size of the enhancement by one power of $|k \eta_0|=k/(a(\eta_0)H)$. Nevertheless as
we have described in Sec.~\ref{sec:threepoint} the remaining effect can feasibly be larger that the usual one. The radiation
transfer functions are dominated by the multipole $l \approx k d_{LSS}$ where $d_{LSS}$ is the distance to the last scattering
surface, and so it is clear that the product $\Delta^T_{l_j}(k_j)\Delta^T_{l_{j+1}}(k_{j+1})\Delta^T_{l_{j+2}}(|k_{j}-k_{j+1}|)$
is dominated by the multipoles which saturate the triangle inequality, namely $l_j=l_{j+1}+l_{j+2}$. The precise form of the
bi-spectrum will depend on our model for the $k$ dependence of the $\beta_{k_i}$, of which we can only guess the form, but the
characteristic peak for the flattened triangles should be sufficient to distinguish this effect from the usual contributions to
non-gaussianity. Note that in the squeezed limit this leading order contribution vanishes, and a subleading term takes over which
is less enhanced. Thus this contribution to the three point function should be clearly distinguishable from the squeezed
triangles contribution.

\section{\label{sec: hiireev}Higher irrelevant operators, N-point functions}

Now let us consider higher order operators and their contributions to various N-point functions. It is possible to make fairly
general statements based on the scaling with $1/a$ as to the magnitude of the terms. For the three point function let us first
consider the effects of higher order operators of the form $\frac{\lambda_n}{M^{2n+4}}\((\nabla \phi)^2\)^n$. Naively since these
are increasingly irrelevant operators we might expect them to scale with increasing powers of $1/a$. However, expanding to cubic
order and defining $\chi=a\delta \phi$ (this is usual since $\chi$ is pure oscillatory at sub-Hubble scales with no scale factor
dependence) we find \be \left .\sqrt{-g} \frac{\lambda_n}{M^{4n-4}}\((\nabla \phi)^2\)^n\right |_{\rm 3-pt}\approx \frac{1}{a^2}
\frac{\lambda_n}{M^{2}} \(\frac{\dot{\phi}}{M^2}\)^{2n-3} (c_1 \chi'^3+c_2 \chi' (\vec{\nabla}\chi)^2). \ee So all the higher
order operators will have the same scaling with $1/a$ and are just additionally suppressed by $\dot{\phi}/M^2$. In particular for
DBI-inflation and k-inflation we will still have an interaction scaling as $1/a^2$ and so similar enhancements to those
described. Similar statements can be made for the N point functions where we find \be \sqrt{-g}
\frac{\lambda_n}{M^{4n-4}}\((\nabla \phi)^2\)^n |_{\rm N-pt}\ \approx \frac{1}{a^{2N-4}} \frac{\lambda_n}{M^{2N-4}}
\(\frac{\dot{\phi}}{M^2}\)^{2n-N} (c_3 \chi'^N+c_4 \chi'^{N-2} (\vec{\nabla}\chi)^2+\dots). \ee To increase powers of $1/a$ we
must go to genuinely higher derivative interactions (rather than powers of first derivatives) such as
$\frac{1}{M^8}\left(\nabla_{a}\nabla_b\phi\nabla^{a}\nabla^{b} \phi\right)^2$. Note that terms containing $\Box \phi$ can be
neglected as they are redundant couplings which can be removed by local field redefinitions. Although the correlation functions
are not invariant under these redefinitions, it is precisely because they are local that they will not give rise to any
interesting effects at late times. Calculating the three point function we have \be \left . \frac{1}{M^8}\sqrt{-g}
\(\nabla_{a}\nabla_b\phi\nabla^{a}\nabla^{b} \phi\)^2\right |_{\rm 3-pt} \approx \frac{1}{M^8a^5} \ddot{\phi}
(\partial_a\partial_b\chi)^2 \chi''\approx \frac{1}{M^8a^5} \ddot{\phi} (\partial_a\partial_b\chi)^2 \partial_i\partial^i\chi.\ee
 So we see that this term
scales as $1/a^5$. Similarly, there are yet higher derivative interactions that we can construct that scale as higher powers of
$1/a$. However, these naive scalings may not directly represent the enhancements that arise due to the cancellation of numerators
evaluated on the flattened triangles. Further more these higher order derivative terms are necessarily additionally $H/M$
suppressed which typically cancels out the extra $1/a$ enhancement. Nevertheless the general form of the enhancement effects for
the three point function implied by effective field theory will take the schematic form (to first order in $\beta_k$) \be
\frac{\Delta \langle \zeta_{\vec{k}_1} \zeta_{\vec{k}_2} \zeta_{\vec{k}_3} \rangle}{\langle \zeta_{\vec{k}_1} \zeta_{\vec{k}_2}
\zeta_{\vec{k}_3} \rangle}=\sum_i |\beta_{k_i}| f\(\frac{k_{t}}{a(\eta_0)M}, \frac{k_i}{k_t},\frac{k_{i+1}}{k_t}\) + {\mathcal
O}(\beta_k^2), \ee where the function $f$ is a dimensionless function which has a Taylor expansion in its first argument and
similar relations holding for the N-pt functions. Thus we see that if $\beta_k \neq 0$, the validity of effective field theory
requires that $k/(M a(\eta_0)) \ll 1$ (for the Taylor expansion to be valid) as we stressed from the outset. With sufficient
e-folds this can be violated and so we discover directly the transplankian problem \cite{transplanckian}, namely the statement
that effective field theory does breaks down at the beginning of inflation if the initial vacuum state is chosen to be
nontrivial. However if $\beta_k=0$ there is no reason to doubt effective field theory even when $k>Ma$. The physical reason is
clear; the BD vacuum contains no particles, so that there is nothing that physically carries transplankian energies, and so no
transplankian effects should be expected.

\section{\label{sec:concl}Conclusions}

Our main conclusions can be summarized as follows: Choosing an excited gaussian (and Hadamard) initial state at the beginning of
inflation can give rise to additional contributions to the bi-spectrum from sub-Hubble interactions. They are significant for
flattened triangles $k_{1}=k_2+k_3$ (and permutations), which in $l$ space saturate the triangle inequality $l_1=l_2+l_3$ (and
permutations). Schematically the corrections from a given operator calculated in an excited state, relative to the equivalent BD
state contribution take the form \be \left . \frac{\Delta \langle \zeta^3 \rangle}{\langle \zeta^3 \rangle}\right |_{\rm
flattened} \sim |\beta_k| |k\eta_0|^{s+1}, \ee where the integer $s$ depends on the precise scaling of the interaction with
inverse scale factor. The specific contribution to $f_{NL}$ for higher derivative interactions is of the order \be \left.
\frac{\Delta \langle \zeta^3 \rangle}{\langle \zeta^3 \rangle}\right |_{\rm flattened} \sim |\beta_k| \(\frac{k}{a(\eta_0)H}\)^2.
\ee The requirement that the excited inflaton quanta do not spoil the slow roll evolution imposes the bound $|\beta_k| <
\sqrt{\epsilon\eta'}\frac{HM_{\rm pl}}{M^2}$ while the validity of effective field theory bounds $\frac{k}{a(\eta)}<M$. Thus the
largest expected enhancement factor is \be \left. \frac{\Delta \langle \zeta^3 \rangle}{\langle \zeta^3 \rangle}\right |_{\rm
flattened} \sim \sqrt{\epsilon \eta'} \frac{HM_{\rm pl}}{M^2} \(\frac{M}{H}\)^2 =\sqrt{\epsilon \eta'} \frac{M_{\rm pl}}{H}\ee which
can easily be of order $10^{4}$. Whilst it may be possible to observe this enhancement in a genuinely three dimensional probe,
the CMB bi-spectrum is only sensitive to a two dimensional projection of this. On converting to $l$ space we must average over
angles which gives a reduction of one factor of $|k\eta_0|$ so that \be \left . \frac{\Delta f_{NL}}{f_{NL}}\right |_{\rm
flattened} \sim |\beta_k| |k\eta_0|^{s}. \ee In particular for the higher derivative interactions, the largest expected value is
\be \left .\Delta f_{NL}\right |_{\rm flattened} \sim \epsilon \sqrt{\epsilon \eta'} \frac{HM_{\rm pl}}{M^2} \frac{M}{H}
\frac{\dot{\phi}^2}{M^4}\sim 100 \sqrt{|\beta_k|} \frac{\dot{\phi}^2}{M^4} \sim 10^4 |\beta_k|^{5/2}.\ee In this particular case
we need $|\beta_k| > 2.5 \times 10^{-2}$ to obtain $f_{NL} \ge 1$. Whether this departure from BD can already be detected in the
power spectrum, depends on the precise $k$ dependence of $|\beta_k|$. Note that in the extreme case in which $\beta_k$ is
constant over the range of scales probed by the CMB, the only affect of $\beta_k$ in the power spectrum would be to renormalize
its amplitude. On the other hand a stronger $k$-dependence could be probed in any spectral tilt present. Nevertheless to get an
enhancement relative to the usual result we only need $|\beta_k| > 10^{-4}$ and so clearly models which already give a large
non-gaussian contribution will be the most interesting in terms of seeing the enhancement.

It is clear from our discussion that typically these effects will only be seen if the number of e-folds of inflation is low, i.e. little more than the require $\sim 55-60$, since we require that the modes corresponding to the observable scales today had physical momenta less that $M \le M_{\rm pl}$ at the onset of inflation. Thus the effect is in the exact opposite region to the transplanckian problem which typically requires $\sim 70$ e-folds for the physical momenta to reach $M_{\rm pl}$. We can imagine a two-stage inflation scenario however where inflation occurs before this period for a possibly large number of e-folds with the usual Bunch-Davies vacuum, and then some non-adiabatic process excites the vacuum state up to scales set by the Hubble constant at that time $\sim M$ (e.g. a sudden change in the trajectory of the inflaton in a multidimensional landscape). Then if a second period of lower scale inflation occurred during which the fluctuations relevant for observable scales today were created, these would arise from this excited state and show the signatures we have described. 

The essential feature of our result was anticipated independently in Ref.~\cite{Chen:2006nt} where a general calculation was performed for theories of the form $p(X,\phi)$. However these authors did not give any estimate as to the possible magnitude of the effect which we have here emphasized is strongly constrained by backreaction. Furthermore as explained there is little chance of observing the full enhancement for the flattened triangles, but rather only smoothed one which is further suppressed by $H/M$. When dealing with theories of the form $p(X,\phi)$ there is an implicit assumption that one can take seriously the effective field theory beyond the normal expected range of validity, by taking seriously the effect of higher derivatives. By contrast in this article our perspective has been that even in standard slow roll minimally coupled inflation, these corrections will exist due to standard effective field theory arguments, and we have remained strictly in a regime in which they are small so that standard effective field theory is valid. To properly analyze the case considered in Ref.~\cite{Chen:2006nt} it will be necessary to redo the backreaction analysis \cite{TolleyHolman}.

Irrelevant operators may have very relevant physical consequences, and indeed as we see here, because the relevant contribution
to the three point function is naturally suppressed, the first irrelevant higher derivative operator can give a larger
correction. For these effects to be observable the cutoff scale of the theory must be less than the Planck scale. Nevertheless,
an effect can easily be present for reasonable scales $H \sim 10^{13-14}\, {\rm GeV}$ and $M \sim 10^{15-16} \, {\rm GeV}$. We
may note that super-horizon approaches to the calculation of non-gaussianities \cite{Salopek:1990jq} will not be able to describe
the effects we have considered precisely because they arise from sub-Hubble scale interactions.

From a theoretical point of view, we can use this calculation to get a better understanding of the special nature of the BD
vacuum as the true vacuum of interacting quantum fields in de Sitter. What we have shown here can be reinterpreted as the
statement that the BD vacuum is self-consistent in that if we choose it to define a Gaussian initial state, although a non-zero
three-point function will be generated, it will be highly suppressed both by slow-roll parameters as well as by factors of $H
\slash M$. Any initial deviations from the BD state will induce potentially large corrections to the three-point function
(although it is important to note that these are still under perturbative control). Keeping the three point function small enough
to be consistent with observations then requires a fine-tuning of the initial state (via the $\beta_k$'s).

We believe these results also offer a new window to address the transplankian problem, since for the reasons we have explained,
higher N-pt functions can be a better probe of the transplankian era that the two point function, provided at least that we start
out in an initially excited state. On the other hand it is likely that if the initial state is taken to be the equivalent of the
BD vacuum in the transplankian region, these effects will be suppressed at least by $H/M$.

In conclusion then we see that the CMB bi-spectrum will be a most useful tool for probing the inflaton initial state, and can
contain richer information than the power spectrum alone. Initial state effects will be present in flattened triangles, and a
direct observation for an enhancement in these triangles would provide the strongest window into the physics at the beginning of
inflation.

\acknowledgments

We would like to thank Latham Boyle, Cliff Burgess, Claudia de Rham, Olivier Dore, Stefan Hofmann, Justin Khoury and Mark Wyman
for useful discussions. R.~H. would like to thank the Perimeter Institute for hospitality during some of this work. He also
acknowledges support from the Department of Energy through DOE grant No.~DE-FG03-91-ER40682. The work of A.~J.~T. at the
Perimeter Institute is supported in part by the Government of Canada through NSERC and by the Province of Ontario through MRI.


\end{document}